\documentclass[english,usenatbib]{mn2e}
\usepackage[]{fontenc}
\usepackage[latin1]{inputenc}
\usepackage{amsmath}
\usepackage{graphicx}
\usepackage{amssymb}
\usepackage[authoryear]{natbib}

\newcommand{\msun}{\mbox{${\rm M}_\odot$}}

\newcommand{\reff}{\mbox{${r_{\rm eff}}$}}
\newcommand{\rc}{\mbox{${r_{\rm c}}$}}

\newcommand{\eq}[1]{\mbox{Eq. #1}}
\newcommand{\fig}[1]{\mbox{Fig. #1}}

\newcommand{\sect}[1]{\mbox{Sect. #1}}

\newcommand{\mean}[1]{\mbox{$\langle #1 \rangle$}}

\title[Mass segregation in star clusters]
      {Mass segregation in young star clusters: can it be detected from the  
integrated photometric properties?}
      \author[Gaburov \& Gieles] {Evghenii Gaburov$^{1,2}$ and Mark Gieles$^3$\\
$^1$ Sterrenkundig Instituut ``Anton Pannekoek'', Kruislaan 403, 1098 SJ Amsterdam, the Netherlands \\
$^2$ Section Computational Science, Kruislaan 403, 1098 SJ Amsterdam, the Netherlands \\
$^3$ European Southern Observatory, Casilla 19001, Santiago, Chile}

\begin{document}
\maketitle

\begin{abstract}
  We consider the effect of mass segregation on the observable
  integrated properties of star clusters. The measurable properties
  depend on a combination of the dynamical age of the cluster and the
  physical age of the stars in the cluster. To investigate all
  possible combinations of these two quantities, we propose an
  analytical model for the mass function of segregated star clusters
  that agrees with the results of N-body simulations, in which any
  combination can be specified.  For a realistic degree of mass
  segregation and a fixed density profile, we find with increasing age
  an increase in the measured core radii and a central surface
  brightness which decreases in all filters more rapidly than what is
  expected from stellar evolution alone. Within a Gyr the measured
  core radius increases by a factor of two and the central surface
  density in all filters of a segregated cluster will be overestimated
  by a similar factor if mass segregation is not taken into account during
  the conversion from light to mass. We find that the $V-I$ colour of
  mass segregated clusters decreases with radius by about 0.1-0.2 mag,
  which could be observable. From recent observations of partially
  resolved extra-galactic clusters, a decreasing half-light radius
  with increasing wavelength was observed, which was attributed to
  mass segregation. These observations can not be reproduced by our
  models. In addition, we provide physical arguments based on the
  evolution of individual stars that one should not expect strong
  dependence of core radius as a function of the wavelength.  We find that the differences between measured radii in
  different filters are always smaller than 5\%.
\end{abstract}

\section{Introduction}\label{sect:introduction}
Observations of star clusters are often used as test-beds for theories
of star formation, the stellar initial mass function (IMF) and
dynamical evolution. An important ingredient in most of these theories
is the differential distribution of stellar masses within a cluster,
or {\it mass segregation}. 

In several resolved clusters, evidence of mass segregation has been
claimed on the basis of observed variations of the stellar mass
function (MF) with distance to the cluster centre (see for example
\citealt{1996ApJ...466..254B,
  1998ApJ...492..540H,2002MNRAS.331..245D,2002A&A...394..459S} for
R136 in 30 Doradus, the Orion Nebula Cluster, clusters in the LMC and
the Arches cluster, respectively). These observations claim an
overabundance of massive stars in the cluster centre. This
stratification of stellar masses is expected from dynamical evolution,
since dynamical friction slows down the most massive stars. As a
result, these stars sink to the cluster centre on a time-scale that is
inversely proportional to their mass. Because the aforementioned
clusters are dynamically young, the observations advocate a primordial
origin for this segregation of stellar masses. 

From the ``competitive
accretion'' star formation model (see for example
\citealt{1997MNRAS.285..201B}), it is expected that the most massive
stars form in the highest density environments, corresponding to the
inner parts of the clusters. The preferential formation of massive
stars in the centre of the cluster is often used to explain the
observations of dynamically young, but mass segregated clusters.

The determination of the stellar MF in different annuli around the
cluster centre, which is the most common technique used to ``detect"
mass segregation in resolved clusters, is hampered by several
observational difficulties. First, crowding and blending of stars in
the core can mimic a shallower MF at that location
\citep{2008ApJ...677.1278M, ascenso}. Second, the determination of stellar masses from the
observed luminosities depends on the adopted age, which is usually
taken constant for all stars in the cluster. However, only a small
spread in age is enough to cause misinterpretations (see
\citealt{1995ApJ...448..179H}, who show that the MF of the
R136 cluster is consistent with Salpeter when this effect is taken
into account).

Alternatively, integrated properties, such as the surface brightness
profile in different filters can be employed to study radial
variations of the stellar MF. In this way one does not have to rely on
individual star counts, thus avoiding possible biases; moreover, this
method can be used for clusters at larger distances. It was shown that
metal-rich (red) star clusters appear to have smaller half-light radii
than their metal-poor (blue) counterparts
\citep{2004ApJ...613L.117J}. In addition, one could expect that
segregated star clusters appear larger in the ultra-violet (UV)
than in the near infra-red (NIR) as most of the light at these red
wavelengths comes from the massive stars, whereas the bluer
wavelengths are dominated by intermediate mass stars. Tentative
evidence for this is given by \citet{2005ApJ...621..278M} who find a
decreasing half-light radius, or effective radius (\reff), with
increasing wavelength for the young massive cluster
M82-F. \citet{2007arXiv0710.0547L} find for NGC~1569-B that the
\reff\ measured in the $U$-band is around 30-50\% larger than the
\reff\ measured in the $I$-band. A smaller radius in the redder
filters is qualitatively what one would expect when the massive (red)
stars are more centrally concentrated, but it has thus-far not been
quantified how large the expected difference is.

Although the integrated properties are free of the biases that are
encountered in methods that rely on individual star counts, there are
other problems such as variations of the PSF between the different
filters \citep{2007arXiv0710.0547L}, differential foreground
extinction \citep{2007MNRAS.379.1333B} and intra-cluster extinction
\citep{2002A&A...394..459S}, that make it a challenging task to
accurately determine intrinsic differences between the cluster
properties in different filters.

To quantify the expected variations in different filters, we present a
method to rapidly simulate integrated luminosity profiles of mass
segregated star cluster in different filters. We choose to do this
analytically to avoid statistical fluctuations which one has to deal
with when considering (realistic) $N$-body systems, and therefore we
do not include the dynamical evolution. Instead, we apply our method
to clusters with different concentrations, which may result from
dynamical evolution.

In \sect{\ref{sect:model}} we introduce the model of the mass function
of a segregated star cluster and in \sect{\ref{sect:observations}} we
present simulated observational properties of such clusters. A discussion and our
conclusions are presented in \sect{\ref{sect:conclusions}}.



\begin{figure*}
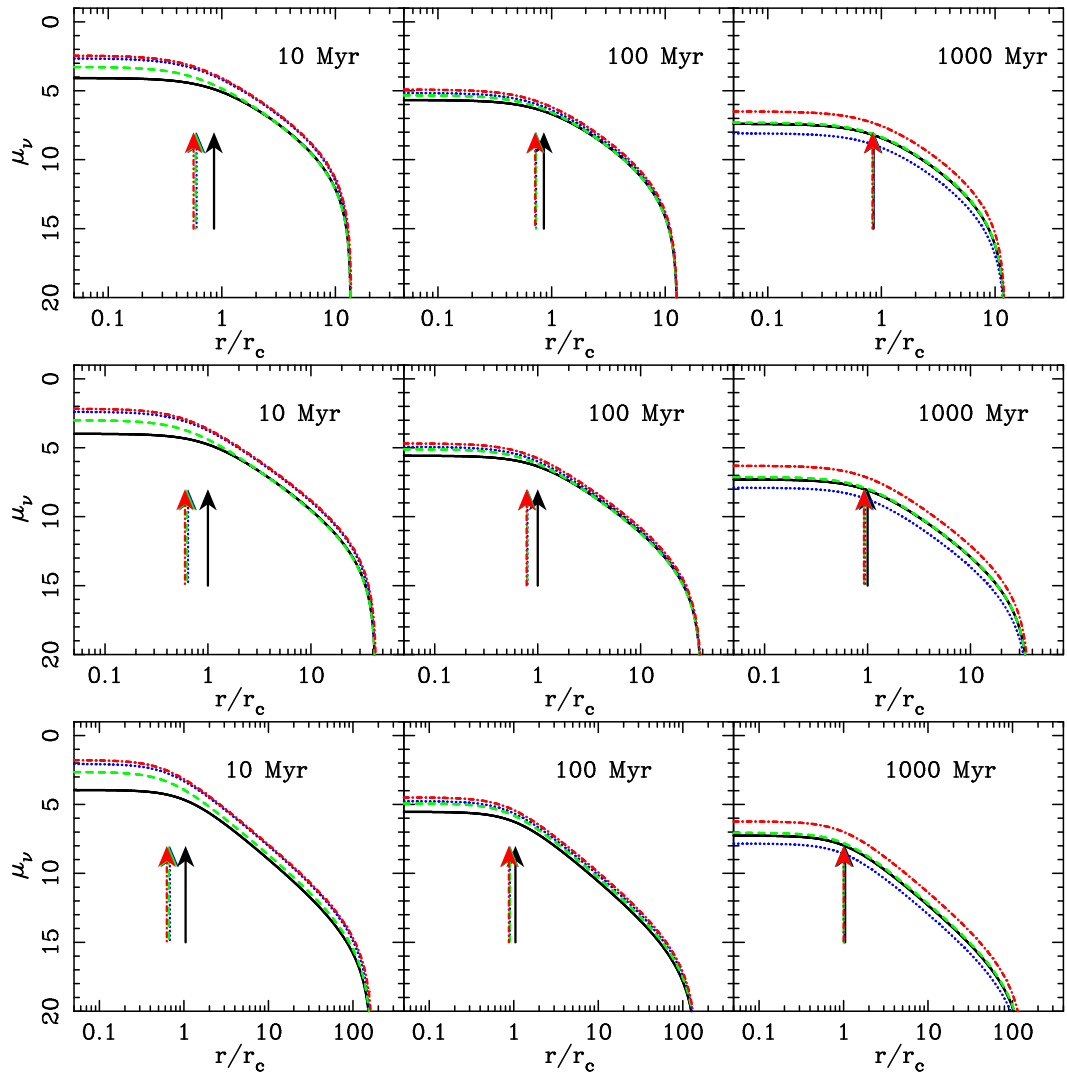

  \begin{center}
    \includegraphics[scale=0.8]{plots/profile_w5.ps}\\
    \includegraphics[scale=0.8]{plots/profile_w7.ps}\\
    \includegraphics[scale=0.8]{plots/profile_w9.ps}\\
  \end{center}
  \caption[]{
    Surface brightness as a function of a distance to the cluster
    centre, which is given in units of core radius of non-segregated
    star cluster. The dotted blue line shows a surface brightness
    profile in $U$-band, dashed green line is for $V$-band and
    dash-dot red line is for $I$-band. The black line represents
    surface brightness profile of a non-segregated cluster in
    $V$-band. The mass density profile for all the clusters of
    different ages is kept the same. The density profiles are King
    models with $W_0=5$, $7$ and $9$ for top, middle and bottom panels,
    respectively.}
  \label{fig:profiles}
\end{figure*}

\section{The Model}\label{sect:model}

%
%
Recent numerical \citep{2007MNRAS.tmpL..45P, 2007arXiv0707.0406G} and
observational \citep{2006ApJ...653L.113K} results suggest that the
mass function in the inner parts of young (dynamically) segregated star clusters has a
broken power-law form with the slope on the high mass end being
shallower.  We assume a segregated mass function (SMF) of the
following form. For $r < r_{\rm hm}$, with $r_{\rm hm}$ being the
cluster's half-mass radius in three dimensions, the SMF is
\begin{equation}
  g(m, r) \propto \left\{
  \begin{array}{ll}
    m^{\alpha_0}, & m_{\rm lo} < m < \mu = 2\mean{m}_0, \\ \mu^{\alpha_0}
    \left({m\over\mu}\right)^{\alpha(r)}, & \mu < m < m_{\rm up}.
  \end{array}
  \right.
  \label{eq:segmf}
\end{equation}
Here $\mean{m}_0$ is the mean mass of the initial MF (IMF), $r$ is the
distance to the cluster centre, $\alpha_0$ is the index of the IMF
which is assumed to be a power-law, $m_{\rm lo}$ and $m_{\rm up}$ are
the lower and upper limits of the MF, respectively, and $\alpha(r)$ is
the $r$ dependent power-law index of the SMF at the high-mass end. The
outer part of the SMF ($g(m, r>r_{\rm hm}) = g_\infty(m)$) does not
depend on distance to the cluster centre. It can be uniquely
constructed in such a way that the integrated cluster mass
function results in the IMF, which we assume to be a
\cite{1955ApJ...121..161S} mass function between $0.15\msun$ and
$100\msun$. In other words, the excess of massive stars in the centre
of the star cluster occurs at the expense of reducing the number of
these stars in the outer regions.



%
%
The form of the mass function depends on the choice of the index
$\alpha(r)$, which can be arbitrary. Guided by $N$-body simulations
\citep{2007MNRAS.tmpL..45P}, we choose the following ``Ansatz''
\begin{equation}
  \alpha(r) = \frac{\alpha_{\rm c} - \alpha_\infty}
        {1 + \left(r\over r_{\rm c}\right)^{3\over2} \frac{\phi}{1 - \phi}}
        + \alpha_\infty.
  \label{eq:ansatz}
\end{equation}
Here $\alpha_{\rm c}$ is a free parameter determining the degree of
mass segregation in the cluster centre, $\phi$ is a free parameter
that specifies the functional form of $\alpha(r)$, $r_{\rm c}$ is the
core radius of the underlying density profile $\rho(r)$, and
$\alpha_\infty$ is a parameter which is constrained by the condition
that the mean stellar mass is a continuous function of $r$, so that
there is no jump at $r_{\rm hm}$.
%
%
For a given $\rho(r)$ there is a maximum possible $\alpha_{\rm c}$
that can be achieved in our model in order to satisfy the constraint
set by the initial MF. We compute this maximum $\alpha_{\rm c}$ by
solving $g_\infty(m_{\rm up}) = 0$. Physically, this means that all
the most massive stars from the outer region are already in the inner
region. Therefore, any further increase in $\alpha_{\rm c}$ will
result in a negative $g_\infty(m)$ for some $m$ below $m_{\rm up}$,
and this is clearly an unphysical situation.

%
%
Given $g(m, r)$ and $\rho(r)$ we can calculate the surface brightness
profile in different broad-band filters. We use the Padova isochrones
for solar metallicity ($Z=0.019$) \citep{1994A&AS..106..275B,
  1996A&AS..117..113G, 2000A&AS..141..371G} and the conversion to the
Johnson-Cousins-Glass $UBVRIJHK$ photometric system
\citep{2002A&A...391..195G} to convert mass to light. Our results do
not depend on the adopted metallicity. Also, we explicitly note that
all stars in the cluster have the same age and metallicity, which
might be an oversimplification, since star cluster with multiple
population are know, both young (for example
\citealt{2008ApJ...681L..17M}) and old (for example
\citealt{2007ApJ...661L..53P}).  The spatial luminosity profile in a
filter centred at wavelength $\lambda$, $L_\lambda(r)$, is computed
using the following conversion $L_\lambda(r) =
\rho(r)[l_\lambda(g(m,r))/\mean{m}(r)]$, in which $\mean{m}(r)=\int
m\, g(m,r)\,dm$ is the mean stellar mass as a function of $r$, and
$l_\lambda(g(m,r)) = \int L_\lambda(m) g(m,r)\,dm$ with $L_\lambda(m)$
being the luminosity of a star of mass $m$ in a filter with central
wavelength $\lambda$.

%
%
We used a series of \cite{1966AJ.....71...64K} models parameterised by
a scaled central potential $W_0$ \citep{1987gady.book.....B} as the
input density profiles; these models were generated by the {\tt starlab}
software package
\citep{2001MNRAS.321..199P}\footnote{http://www.manybody.org/starlab}. We
obtain the projected core radius and the surface brightness profile
by fitting a \citet{1962AJ.....67..471K} profile to our models after
projecting them in 2D. 


%
%
Currently, our model contains three free parameters: $W_0$,
$\alpha_{\rm c}$, $\phi$. However, by considering only the maximal
value of $\alpha_{\rm c}$, we reduce the number of free parameters to
two because the index $\alpha_{\rm c}$ is constrained by $W_0$. In
this case, the value of $\alpha_{\rm c}$ corresponds to the maximum
degree of segregation which is reached at the moment of core collapse
and remains roughly constant after that \citep{2007MNRAS.tmpL..45P,
  2007ApJ...655L..45M}.  Our results can therefore be considered as
upper limits, since smaller values for $\alpha_{\rm c}$ will weaken
the imprint of mass segregation on the integrated properties.  We
focus our studies on King models with $W_0 = 5, 7, 9$. We find that
the simulated properties depend weakly on the value of $W_0$. Fitting
\eq{\ref{eq:ansatz}} to $N$-body simulations
\citep{2007MNRAS.tmpL..45P}, we find that $\phi\simeq
0.3$. Nevertheless, our numerical experiments show weak dependence of
the integrated properties on this parameter. Thus, we generally
present our results for $\phi = 0.5$ unless mentioned otherwise.

Even though there are other methods which include mass segregation in
star cluster models \citep{2008MNRAS.385.1673S, 2008MNRAS.386.2047M},
we prefer this broken power-law approximation because it was found in
dynamical simulations. In addition, we also provide physical
argumentation based on stellar evolution why our results are not
expected to be sensitive to stellar metallicity and the details
of the model of mass segregation.

\section{Results}\label{sect:observations}
Our aim is to understand the effect of mass segregation on the observed size of
clusters, particularly on $r_{\rm c}$ and the surface brightness
profiles in different filters, as a function of age. In addition, we
study the radial variation of colour, since it is expected that the
central part of the cluster is redder than the outer parts, due to the
overabundance of massive (red) post main-sequence stars there
\citep{2006MNRAS.369.1392F}. We note that we allow the stars to evolve
while fixing the density profile, which is of course not realistic
from a dynamical point of view. In reality, the measurable properties
of mass segregated clusters will result from a complex interplay
between mass loss by stellar evolution and dynamical relaxation
processes. Such models only allow the choice of one combination of
dynamical and physical age making the exploration of the full
parameters space too time-consuming for the scope of this study. We
refer the reader to the studies of \citet{2007MNRAS.379...93H} and
\citet{2007MNRAS.379L..40M} as examples of full $N$-body studies that
take into account stellar evolution and the effects of projection.


\subsection{Core radii}

Mass segregated star clusters have an excess of massive stars in their
central regions. As a result, they are expected to appear smaller in
the NIR than in UV \citep{1998ApJ...506..721S}, since the light at
these wavelengths is dominated by massive stars. If this is correct,
it may provide a robust method for determining mass segregation in
slightly resolved (extra-galactic) star clusters at distances up to
several Mpc.

In \fig{\ref{fig:profiles}} we show for different ages the surface
brightness profiles for mass segregated clusters with different
density profiles in the $U$, $V$ and $I$ filters resulting from our
model (Sect.~\ref{sect:model}). We use the same $\rho(r)$ at all ages
in order to eliminate effects related to the dynamical evolution of
the cluster.


At an age of 10\,Myr
the difference in \rc\ between the $U$ and $I$ filter is just 6\%,
with the smallest radius in the $I$-band. The difference between the
true and observed core radius is roughly a factor of two, with the
measured radius being smaller. This is because the massive stars
dominate the light in all filters and are overabundant in the centre.
In the course of time, massive stars leave the main-sequence and
become dark objects, such as black holes or neutron stars, which
results in an apparent increase of the core radius. The observed core
radii become roughly 80\% of the true core radii at this age and the
difference between core radii in the different filters has nearly
disappeared.  Finally, at 1\,Gyr the light is dominated
by red-giant and AGB stars. The turn-off point is close to
2\,M$_\odot$ and the light-to-mass ratio in the core is close to
unity. We find that the observed core radius is 10\% smaller than the
true core radius, thus providing a good estimate of the true
\rc. However, it is unclear whether at 1\,Gyr the mass function can
still be represented by a broken power-law (\eq{\ref{eq:segmf}}). In
fact, recent results suggest that in globular clusters, it is the
low-mass part of mass function which becomes $r$ dependent
\citep{2000ApJ...530..342D,2003MNRAS.340..227B}.  We note that in our
model we have assumed that the mass of all remnants is retained, that
is, we do not take into account kick velocities, and the mass of the
remnant we estimate from its zero age main sequence mass
\citep{1989ApJ...347..998E, 1996A&A...309..179P}


The {\it apparent} increase of \rc\ with age implies that light is not
a good tracer of mass for young mass segregated clusters.
\citet{2003MNRAS.338...85M} observed a clear trend of increasing core
radius with age, which can partially be explained by the effect of
mass segregation making the younger clusters appear more compact.
However, dynamical effects, such as heating by black hole binaries
\citep{2004ApJ...608L..25M, 2007MNRAS.379L..40M} are still required
to fully explain the observed trend of the increasing \rc\ with age for the LMC clusters.

\begin{figure*}
  \begin{center}
    \includegraphics[scale=0.3]{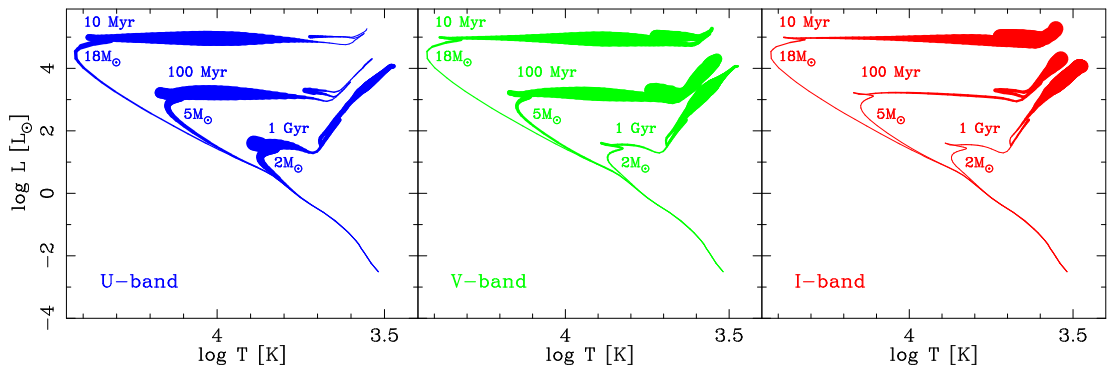}
  \end{center}
  \caption[]{Hertzsprung-Russell diagram of stellar isochrones in
    different filters. The thickness of the line represents the contribution
    to the cumulative luminosity function, that is $L_\lambda(m)p(m)$
    in which $p(m)$ is the initial mass function.}
  \label{fig:hrdiag}
\end{figure*}

\begin{figure*}
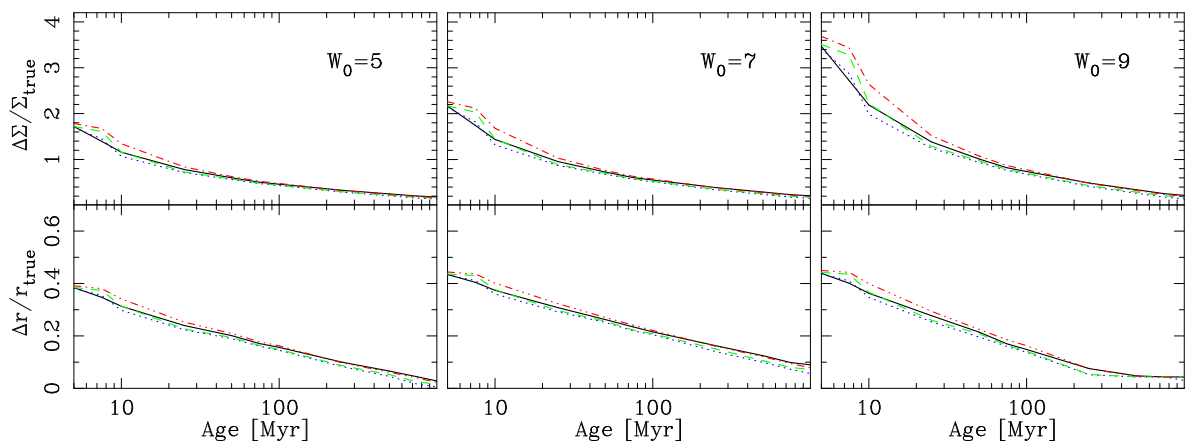

  \begin{center}
    \includegraphics[scale=0.29]{plots/density_w5.ps}\vspace{-1pt}
    \includegraphics[scale=0.29]{plots/density_w7.ps}
    \includegraphics[scale=0.29]{plots/density_w9.ps}
  \end{center}
  \caption[]{Upper panel: The overestimate of the observed central
    surface density, $\Delta\Sigma = \Sigma_{\rm obs} - \Sigma_{\rm
      true}$, compared to the true central surface density in mass
    segregated star clusters as a function of age. The solid black
    line displays the overestimate assuming bolometric light-to-mass
    ratio (dotted blue line is for light-to-mass ratio in $U$ filter,
    dashed green line is for $V$ filter, and dash-dotted line is for
    $I$ filter). Lower panel: The underestimate of the observed core
    radius, $\Delta r = r_{\rm true} - r_{\rm obs}$, compared to the
    true core radii as function of age. The notation is the same as in
    the upper panel. The density profiles are King models with
    $W_0=5$, $7$ and $9$ for left, middle and right panels.}
  \label{fig:density}
\end{figure*}


In the lower panel of \fig{\ref{fig:density}}, we show the time
evolution of the difference between the true \rc\ and the observed
\rc\ in the $U$, $V$ and $I$ filters, as well as the \rc\ computed
from the bolometric surface brightness profile. As we have shown
before, the difference between the true and the observed
\rc\ decreases in the course of time from roughly 50\% at 10 Myr to
about 10\% at 1 Gyr. This result is quite robust showing only a weak
dependence on the choice of the parameters $W_0$ and $\phi$.

The reason that the core radii are similar at all wavelengths can be
understood from some simple arguments. The light in all filters is
dominated by stars with masses slightly above the turn-off mass
(\fig{\ref{fig:hrdiag}}). The red filters are dominated by giant stars
at all ages. The optical and blue filters are also dominated by stars
that are close to the turn-off point and not by the stars on the
main-sequence. This is because the increase of the light-to-mass ratio
with mass is much stronger than the decrease of the number of stars,
and this holds for any realistic MF. That is, stars of similar
mass dominate the light in all the filters at all ages. However, since
stars of similar mass have similar spatial distribution in a mass
segregated star cluster, the differences in the observed core radii
are small. In the case of mass segregation, MF is a function of
  the distance to the cluster centre, $r$, such that the mean stellar
  mass decreases with $r$. Since most of the light comes from stars of
  similar mass, independent of the shape of the MF, the observed core
  radii are weakly dependent on how mass segregation is implemented in
  our model.


The light-to-mass ratio in the centre of a segregated cluster can be
several times larger than that of a simple stellar population (SSP) of
the same age. Usually, the central {\it mass} density of a star
cluster is estimated by using the light-to-mass ratio of an SSP model
corresponding to the age of the cluster to convert the surface
brightness profile to a surface density profile. Since in segregated
clusters the light-to-mass ratio in the centre is higher, the central
density will be overestimated. If the observed central surface
brightness is $\mu$, one can compute the ``observed'' central surface
density, $\Sigma_{\rm obs} = (M/L)_{\rm ssp} \mu$. The true surface
density in the centre, $\Sigma_{\rm true}$, we obtain from the input
density profile. The relative difference between these two quantities
is displayed in \fig{\ref{fig:density}}.  Deviations from one are most
prominent in the first 100 Myr, where the maximum difference is about
a factor of two. This is not a sensitive function of $\phi$. However,
in more concentrated clusters, $W_0=9$, the density overestimation is
larger by an additional factor of two (\fig{\ref{fig:density}})

Our findings could explain the observed trend of decreasing central
surface density with age in a sample of LMC and SMC clusters
\citep{2007AJ....134..912N}. If star clusters are born
with a high central concentration, their central  3D density will be overestimated by
nearly an order of magnitude.

\subsection{Colour gradients}

An additional tracer of mass segregation is the colour of a star
cluster as a function of the distance to the cluster centre. Using a
gas dynamics code, \citet{2006MNRAS.369.1392F} studied mass
segregation in young star clusters and found that the $V-K$ difference
between the inner and outer part of the cluster is roughly 0.1
magnitude in the first few 10\,Myr, with the inner parts being
redder. This difference, though small, might be observable.

\begin{figure*}
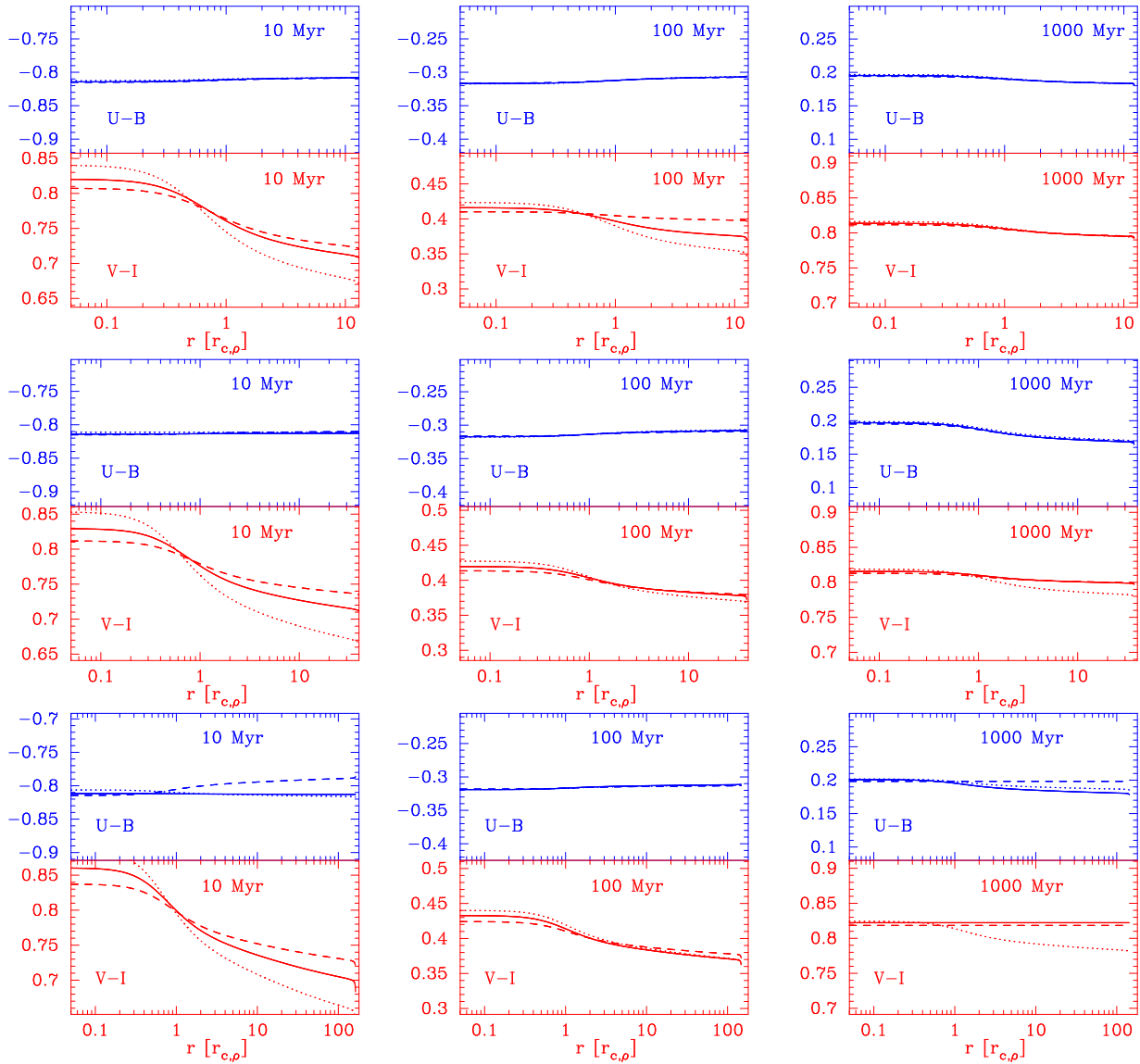

  \begin{center}
    \includegraphics[scale=0.33]{plots/colours_w5.ps}
    \includegraphics[scale=0.33]{plots/colours_w7.ps}
    \includegraphics[scale=0.33]{plots/colours_w9.ps}
  \end{center}
  \caption[]{Colour of the cluster as a function of projected distance
    to the cluster centre. The upper three panel shows $U-B$ colour,
    whereas the lower panel displays $V-I$ colour. The solid line
    represent displays colours of the cluster if $\phi = 0.5$, dashed
    line for $\phi=0.3$ and dotted line for $\phi=0.7$. The top panels
    use King density profile with $W_0=5$, while the middle and bottom
    panels use King profile with $W_0=7$ and $9$ respectively.  
  }
  \label{fig:colours}
\end{figure*}

In \fig{\ref{fig:colours}} we display the colour as a function of $r$
at different ages resulting from our models.  The colour gradients has
a notable dependence on $\phi$ and, therefore, we covered three
different values of $\phi$: 0.3, 0.5, and 0.7. Note that these three
cases are all for the maximum degree of mass segregation, that is set
by $\alpha_c$. The value of $\phi$ controls the change of the MF with
$r$ and the colour variations are sensitive to the value of $\phi$. In
all cases, the largest colour gradient is observed only for the
clusters with ages younger than a few times 10\,Myrs.  The largest
colour gradient of $V-I$ is roughly 0.2 magnitude, whereas the
gradient of $U-B$ is barely observable. The clusters with high
concentration exhibit higher colour gradient than the cluster with low
concentration. In the course of time, however, the colour gradients
become less prominent. We, therefore, expect that mass segregation might
be detectable in young ($<100\,$Myr) star clusters through radial
colour variations in $V-I$ or $V-K$.


\section{Discussions and Conclusions}\label{sect:conclusions}
We have simulated observational properties of mass segregated star
clusters with the aim to quantify the imprint of mass segregation on
integrated cluster properties. We choose to model the segregated mass
functions analytically based on results from observations and $N$-body
simulations.  In order to calculate upper limits of the imprint of
mass segregation, we only considered clusters with the maximum
possible degree of segregation which can be achieved in our
model. While this may not necessary be reached through dynamical
evolution, one may think of this setup as the result of a combination
of primordial segregation combined with dynamical evolution.

In young ($\sim10\,$Myr) star clusters, we find only small differences
($\lesssim5\%$) between the core radius (\rc) found in different
filters, and the differences becomes smaller for older clusters. The
explanation for these small differences is that the most massive stars
dominate the light in all filters at all ages. It, therefore, appears
that the comparison of the measure \rc\ in different filters is not a particularly suitable tool to look for 
mass segregation in star clusters and, therefore, we do not expect that
one could study whether the young star cluster is in core-collapse
phase by using the light alone
\citep{2006ApJ...653L.113K,2007MNRAS.tmpL..45P}.

The observed \rc\ is underestimated for young segregated star
clusters. The difference with the real \rc\ decreases with age and is
only 10\% at 1 Gyr compared to nearly a factor of two at 10 Myr and
this does not depend on the cluster concentration.  The same factor
was found for the half-light radius in gaseous models of mass
segregated clusters by \cite{2005ApJ...620L..27B}.

The underestimation of \rc\ results in an overestimation of the
observed central density. This is because the light-to-mass ratio in
the centre of a segregated cluster can be an order of magnitude higher
than that of a simple stellar population without segregation and,
therefore, central 3D mass density might be overestimated by nearly an
order of magnitude. However, due to the projection the measured
central surface density is overestimated by a factor of three, and
this result slightly depends on the choice of $W_0$. In particular,
for clusters with high concentration the central density might be
overestimated by nearly an order of magnitude. These results are
consistent with the finding of \cite{2005ApJ...620L..27B}.

The observed central surface and core radius approached its true value
around 1 Gyr. We therefore conclude that light is not a good tracer of
mass in young ($<100\,$Myr) star clusters that are segregated and this
effect should be taken into account when trends of \rc\ or central
surface density with age are discussed \citep{2007A&A...469..925S,
  2007AJ....134..912N, bastian08}. For clusters with ages above 100\,Myr these
differences are smaller than 20\%.

It is possible to observe mass segregation by looking at colour
differences between the inner and outer parts of star clusters. We
find that our simulated clusters have $V-I$ differences of roughly
0.1-0.2 magnitude between the centre and the outer part of the
cluster. This effect has been observed by \cite{2007arXiv0710.0547L}
in the $\sim20\,$Myr old massive cluster NGC~1569-B. The colour
difference decreases at older ages.


We would like to stress that our analytic results for the surface
  brightness profiles represent an observational best case
  scenario. From the observational data, the accuracy of the
  \rc\ determination will be limited by photon noise, and perhaps more
  importantly, by stochastic effects in the stellar IMF. Several
  studies have quantified the effect of stochastic fluctuations (see
  for example \citealt{2000ASPC..211...34L, 2002IAUS..207..616B,
    2004A&A...413..145C}). \citet{2000ASPC..211...34L} determine the
  age dependent minimum mass a cluster should have, such that the
  relative fluctuations around the mean flux, $\sigma_L/L$, is less
  than 10\%, corresponding to roughly 0.1 mag photometric
  uncertainty\footnote{Photometric uncertainty is calculated form the
    flux uncertainty in the following way, $\sigma_V = -2.5\log(1 +
    \sigma_L/L)$ mag.}. When flux is determined in the $V$-band, this
  minimum mass at ages of [10, 50, 200, 1000]\,Myr is [$10^5, 2\,10^4,
    10^4, 6\,10^3$]\,$\msun$.
  
  We predict that the most prominent feature of mass segregation in
  the integrated properties is a 0.1 mag difference in $V-I$ colour
  (Fig.~\ref{fig:colours}) between the inner part and the outer
  part. In order to be able to report a detection of this difference,
  one needs an uncertainty of $\sigma_{V-I} <<0.1$ mag and we adopt  $\sigma_L/L \simeq 0.01$.


\citet{2000ASPC..211...34L} show that $\sigma_L/L$ scales with
$(\sqrt{N}L)^{-1} \propto N^{-3/2}$, since $L \propto N$. This implies
that the minimum masses quoted above have to be a factor
$(0.1/0.01)^{2/3}\simeq4.6$ higher to be able to distinguish a radial
colour variations due to mass segregation from stochastic fluctuations
due to IMF sampling. For the 10\,Myr case this implies a minimum mass
of $\sim 5 \times10^5\,\msun$. This means it will be difficult to
detect mass segregation in young massive Galactic clusters, such as
the Arches clusters or Westerlund 1. Even the cluster R136 in the 30
Doradus region in the Large Magellanic Cloud is probably of too low
mass to detect mass segregation. R136 is probably a special case
anyway, since it does not have red evolved stars, which all of our
models do have. Young clusters more massive than $\sim 5
\times10^5\,\msun$ are known, for example in the Antennae galaxies
\citep{1995AJ....109..960W} and M51 \citep{2000MNRAS.319..893L,
  2005A&A...431..905B}, but these are too distant ($\gtrsim10\,$Mpc)
to be able to determine a colour gradient in the light profile.  There
are probably only a handful of candidate clusters that are resolved
enough such that a colour gradient can be observed (for example some
of the young massive cluster in M82 \citep{1995ApJ...446L...1O}, the
massive ``young globular cluster" in NGC~6946
\citep{1967PASP...79...29H, 2002ApJ...567..896L} and a few clusters in
M83 \citep{2004A&A...427..495L}.
 
An alternative method to detect differences in the IMF between the
inner and outer parts would be to determine the spectral
properties. If this can be done without changing instrument set-up,
then no problems with changing PSFs or weather conditions should
affect the observations. It would, therefore, be interesting to
investigate which spectral range would be most sensitive to changes of
the slope in the IMF.

\section{Acknowledgements}
We thank Nate Bastian for ideas and suggestions and Simon Portegies
Zwart, Henny Lammers, S{\o}ren Larsen and Andres Jord{\'a}n for
helpful discussion. EG is supported by NWO under the grant
\#635.000.303.

\bibliographystyle{mn2e} \bibliography{GG2007}

\end{document}